# Topographic forcing of submesoscale instability in the Antarctic Circumpolar Current

Laur Ferris[a], Donglai Gong[a], John Klinck[b],

[a]*Virginia Institute of Marine Science - William & Mary, Gloucester Point, VA 23062, USA*
[b]*Center for Coastal Physical Oceanography - Old Dominion University, Norfolk, VA 23529 USA*

Subpolar frontal zones are characterized by energetic storms, intense seasonal cycles, and close connectivity with surrounding continental shelf topography. At the same time, predicting the ocean state depends on appropriate partition of resolved and parameterized dynamics, the latter of which requires understanding the dynamical processes generating diffusivity throughout the water column. While submesoscale frontal instabilities are shown to produce turbulent kinetic energy (TKE) and mixing in the surface boundary layer (SBL) of the global ocean, their development in complex dynamical regimes (e.g., elevated preexisting turbulence, large ageostrophic shear, or in proximity to topographic boundaries) is less understood. This study investigates the development of submesoscale instabilities, i.e. symmetric instability (SI) and centrifugal instability (CI), near topographic boundaries using a hindcast model of the Drake Passage and Scotia Sea region. The model suggests subsurface SI and CI are widespread along the northern continental margins of the Antarctic Circumpolar Current (ACC) due to topographic shearing of the anticyclonic side of Polar Front jets. Forced instabilities may facilitate persistent mixing along Namuncurá - Burwood Bank, as well as in other southern (northern) hemisphere currents with low potential vorticity and a seamount or sloping topography on the left- (right-) downstream side.

*Email address for correspondence: lnferris@vims.edu*

## Introduction

Mesoscale processes and turbulent mixing within the Southern Ocean play critical roles in global circulation and climate; but their exact relationship, including the relative importance of isopycnal and diapycnal processes, is still poorly understood (Waterhouse et al., 2014; Tamsitt et al., 2017). The Southern Ocean is characterized by filament-like density fronts, water mass boundaries demarcated by abrupt changes in the temperature-salinity relation which give rise to strong zonal geostrophic jets and sites of concentrated mesoscale eddying. The positions of ACC fronts and their associated geostrophic flow, eddy kinetic energy, poleward heat flux, and carbon uptake vary on seasonal to inter-annual timescales, responding to forcing changes such as the Southern Annular Mode and climatological warming (Meredith & Hogg, 2006; Lenton & Matear, 2007; Liau & Chao, 2017).

The dynamics of intense frontal regions are challenging for numerical ocean models to predict, both on operational and climatological timescales. One issue is that Reynolds Averaged Navier-Stokes (RANS)-type ocean models implement mixing through diffusivity parameterizations (i.e. KPP [Large et al., 1994], Mellor-Yamada [1982], Generic Length Scale [GLS]) generally based on vertical buoyancy and velocity gradients. Submesoscale instabilities such as SI and CI arise in part from horizontal buoyancy and velocity gradients; the unresolved mixing effects of these instabilities are not represented by traditional subgrid-scale mixing parameterizations. Additionally, some parameterizations separate physical processes into surface effects and interior effects. For example, KPP leverages boundary layer similarity scaling in the upper ocean and three interior processes (shear instability, double diffusive mixing, and internal waves). Regions where energetic currents flow through complex topography can fall outside the design conditions of these parameterizations; boundaries are known to alter stability (e.g., Gula et al., 2016; Yankovsky et al., 2021), allowing traditionally surface-based instabilities to occur in the ocean interior. With limited computational resources, it is advantageous to identify the processes most influential to mixing; and whether to parameterize these instabilities as SBL processes or throughout the interior in the development of next-generation climate and regional models.

SI has gained interest for explaining enhanced mixing at frontal jets; and arises from the same physical setup as baroclinic instability, but acts at smaller scale and in the across-front direction (Smyth & Carpenter, 2019). CI (Jiao & Dewar, 2015) occurs when absolute vorticity destabilizes the flow, independent of any destabilization by density effects. An inviscid criterion for SI in a steady geostrophic flow is $Ri < f/\zeta_a = f/(f + V_x - U_y)$, termed centrifugal-symmetric instability (CSI) when relative vorticity has a significant-but-insufficient role in destabilization. Notably Chor et al. (2022) used large eddy simulations to find that CSI carries a higher mixing efficiency (the fraction of TKE that meaningfully alters the water column) than SI, indicating this less-idealized variety of submesoscale instability may play a disproportionate role in mixing some regions of the global ocean. Furthermore, external forcing can sustain SI despite its removal by shear production, buoyancy production[1], and dissipative processes. Ekman buoyancy flux (EBF), created when along-front wind stress causes an Ekman advective transport of dense water over light water, is one type forcing that can sustain baroclinic and symmetric instability[2]. However, an along-front wind component is not required for SI; it is one of many agents in reducing (enlarging) the potential vorticity. The notable agent in this paper is topographic drag.

We take a moment to highlight two distinct diffusivity parameterizations for SI in the literature: Bachman et al. (2017) which considers surface-forced SI, and Yankovsky et al. (2021) which considers SI throughout the water column. The Bachman parameterization, applied to the Coastal and Regional Ocean Community Model (CROCO) by Dong et al. (2021), treats geostrophic production by forced SI (FSI), which can occur when EBF and the surface buoyancy flux ($J_b$) sustain SI in the SBL, $EBF + J_b > 0$. Here $EBF = |\tau U_z \rho_0^{-1}| \cos\theta_w$, where $\tau$ is the wind stress, and $\theta_w$ is the angle of the wind relative to geostrophic shear ($U_z$). The parameterization (Bachman et al., 2017; Dong et al., 2021) uses the bulk potential vorticity

$$q = (f\hat{k} + \nabla \times u) \cdot \nabla b = \Big[B_z(f + V_x - U_y)\Big]_{vertical} + \Big[B_y U_z - B_x V_z\Big]_{lateral} \quad (1)$$

to identify instability ($qf < 0$) in the SBL and estimates the associated geostrophic shear production from FSI. In contrast, Yankovsky et al. (2021) developed a parameterization for SI throughout the water column which does not rely on dimensional parameters or FSI. Our results suggest SI below the SBL is ubiquitous in topographically sheared frontal regions, indicating that subsurface parameterization (e.g., Yankovsky et al., 2017) is favored in regions with complex topography.

---

[1] Wienkers et al. (2021) examines the ratio of shear:buoyancy production as a function of front strength.

[2] This effect can be augmented by nonlinear Ekman dynamics acting on a nonuniform vorticity field, such as a jet with an cyclonic side and an anticyclonic side (Thomas et al., 2008).

To best parameterize submesoscale instabilities, an improved understanding of their phenomenology is needed. Several prior efforts have aimed to elucidate the role of submesoscale dynamics in complex regimes. Gula et al. (2016) used a nested Regional Ocean Modeling System (ROMS) model ($\Delta x = 200$ m) to show that the anticyclonic (eastern) side of the Gulf Stream is topographically sheared by the Bahama Banks, decreasing relative vorticity (amplifying anticyclonic shear) sufficient to produce CI. Dewar et al. (2015) discuss a similar mechanism in the smaller-scale, estimating diffusivities of $10^{-4}$ $m^2/s$ due to topographically forced CI. St. Laurent et al. (2019) used a HYCOM model ($\Delta x = 1/12°$) of Palau's wake and a turbulence glider to show only 10% of elevated TKE is attributable to classic wind-driven shear --- the other 90% of elevated TKE likely attributable to shear or submesoscale instability associate with the relative vorticity field[3] in Palau's wake. Rosso et al. (2015) used a hydrostatic MITgcm model ($\Delta x = 1/80°$ or ~1.39 km) to study forward energy cascade in the south Indian ACC, suggesting mesoscale EKE and strain rate could be used to parameterize submesoscale vertical velocity. Mashayek et al. (2017) used nested 1/100º model to show topographic enhancement of mixing over various hotspots in the Drake Passage and Scotia Sea, again confirming the strong role of topography in the ACC forward energy cascade. Finally, Wenegrat et al. (2018; 2020) discuss the importance of baroclinic instability, CI, and SI in the Ekman adjustment of bottom boundary layers over sloping topography.

Observations of the Kuroshio (D'Asaro et al., 2011) and Gulf Stream (Thomas et al., 2013; Thomas et al., 2016; Todd et al., 2016) have suggested SI might be ubiquitous to the ACC. There are limited observations of symmetric instability in the ACC, but one instance is Adams et al. (2017), who observed a variety of submesoscale instabilities in the upper 200 m of a mesoscale cyclonic eddy in the Scotia Sea during the SMILES (Surface Mixed Layer Evolution at Submesoscales) project. Submesoscale instabilities resulting from the interaction of mesoscale eddies with the Polar Front (PF) were shown to generate large vertical velocities (~100 m/day) and water mass modification associated with the Sub Antarctic Mode Water (SAMW). The largest patch of CI was on the edge of a warm core ring closest to sloping bathymetry at 100-150 m depth, with other areas dominated by gravitational, symmetric, and mixed instabilities. The northern edge of the eddy was within 1/2º of the North Scotia Ridge, such that it is compelling to consider whether topography influenced these instabilities. Naveira Garabato et al. (2019) used a microstructure-equipped AUV in an along-slope current of South Orkney Plateau, Antarctica, to observe that submesoscale instabilities (including SI) drive a cross-current secondary circulation and expedite the transformation of water through enhanced boundary layer-interior exchange. It seems the flanks of geostrophic currents, where horizontal shears are greatest, may be more active sources of submesoscale instability than the geostrophic fronts themselves.

[3] Simmons et al., (2019) discuss how vorticity structures in the wake draw energy from mean flow and feed energy to smaller scales, where instability converts this energy to TKE dissipation.

A November 2017 - February 2018 glider program, Autonomous Sampling of Southern Ocean Mixing (AUSSOM), also measured moderately elevated TKE dissipation rates where the ACC flows past Namuncurá - Burwood Bank (Fig. 1). This turbulence record shows elevated turbulence in three distinct regimes: the SBL, the subsurface ocean near the Bank (along both the continental rise and over the shelf), and subsurface open ocean (1-12 December). For further details about AUSSOM the reader is referred to Ferris (2022) and Ferris et al. (2022a; 2022b), and for further details about glider-based turbulence measurement, the reader is referred to Fer et al. (2014) and St. Laurent & Merrifield (2017). In this study, we use vorticity and buoyancy flux fields from a 1-km ROMS hindcast (developed in support of AUSSOM) to show that topographic shearing drives CI and CSI when the PF veers close to the northern boundary of the ACC in the Drake Passage and Scotia Sea[4], providing one mechanism for elevated turbulence along Namuncurá - Burwood Bank. More generally, this represents a pathway for submesoscale frontal instabilities to supply energy to the ACC microscale.

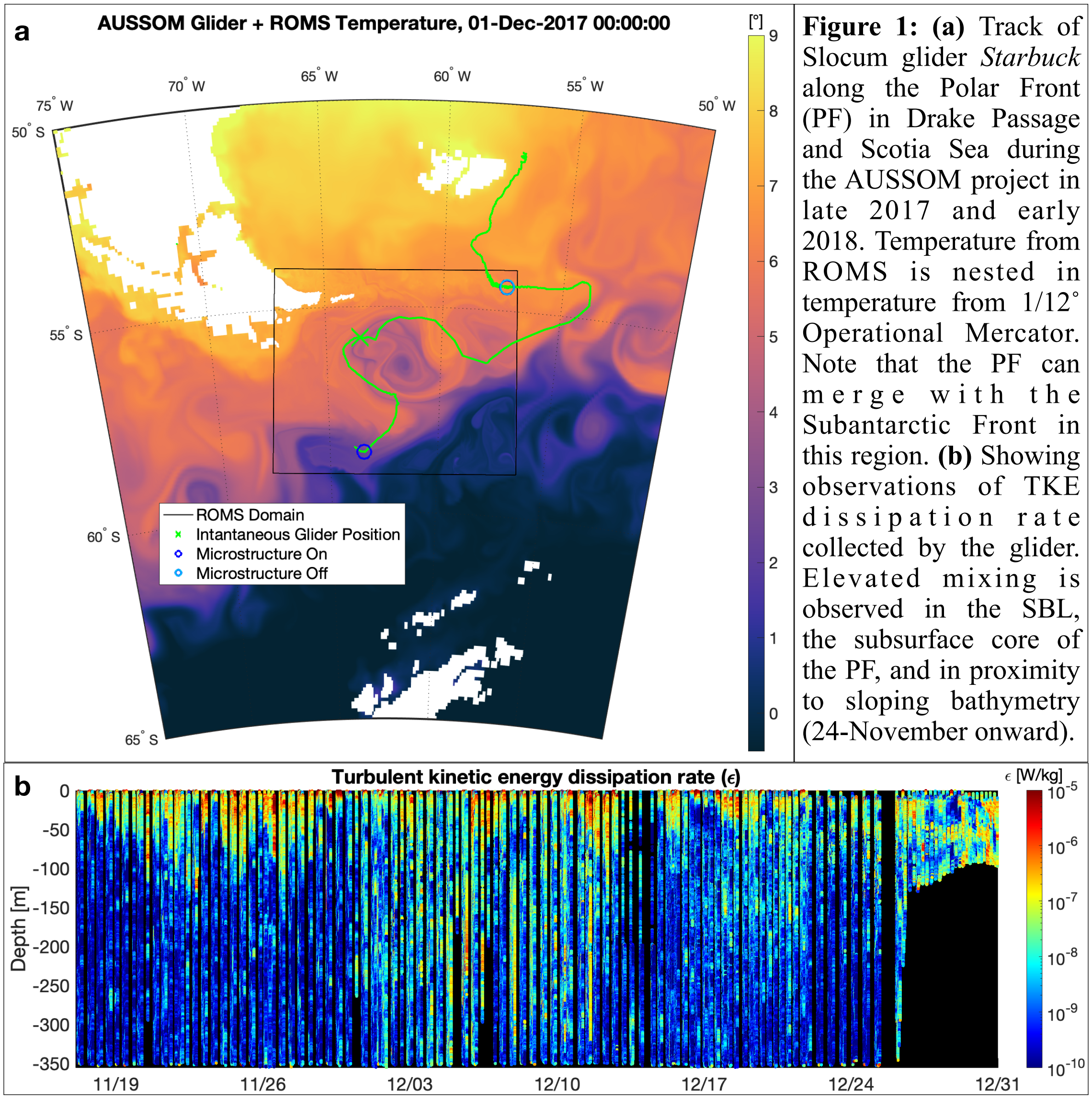

**Figure 1: (a)** Track of Slocum glider *Starbuck* along the Polar Front (PF) in Drake Passage and Scotia Sea during the AUSSOM project in late 2017 and early 2018. Temperature from ROMS is nested in temperature from 1/12° Operational Mercator. Note that the PF can merge with the Subantarctic Front in this region. **(b)** Showing observations of TKE dissipation rate collected by the glider. Elevated mixing is observed in the SBL, the subsurface core of the PF, and in proximity to sloping bathymetry (24-November onward).

**Methods**

For this basin-scale analysis, we use a criterion (Eq. 2) for overturning instability (Hoskins, 1974; Thomas et al., 2013) based on the balanced Richardson number $Ri = B_z/(U_z^2 + V_z^2)$, noting it assumes the dominance of geostrophic dynamics, $B_y = -fU_z$.

$$Ri_B = \frac{N^2}{U_z^{G2} + V_z^{G2}} \equiv \frac{f^2N^2}{|\nabla_h b|^2} < \frac{f}{\zeta_a} \tag{2}$$

Excluding barotropic CI, $f\zeta_a < 0$, overturning instabilities arise when $\Phi_{Ri_B} < \Phi_c$, where $\Phi_{Ri_B} = \tan^{-1}(-1/Ri_B)$ and $\Phi_c = \tan^{-1}(-\zeta_a/f)$. Here $\zeta_a$ is the absolute vorticity and $b = -g\rho_\theta/\rho_0$. The inverse tangent function can be approximated as a piecewise function such that discretized instability types are identified by the relative dominance of terms, which is useful for the compact identification of instability types (Table 1; Thomas et al., 2013).

Table 1. Instability Criteria

| Type | Criteria |
|---|---|
| Centrifugal Instability (CI) | $f\zeta_a < 0$ and $B_z > 0$ |
| Gravitational Instability (GI) | $-180 < \Phi_{RiB} < -135$ |
| Gravitational-Symmetric Instability (GSI) | $-135 < \Phi_{RiB} < -90$ |
| Symmetric Instability (SI) | $-90 < \Phi_{RiB} < \Phi_c$ and $\Phi_c < -45$<br>or<br>$-90 < \Phi_{RiB} < -45$ and $-45 < \Phi_c$ |
| Centrifugal-Symmetric Instability (CSI) | $-45 < \Phi_{RiB} < \Phi_c$ and $-45 < \Phi_c$ |

Model output with 1-km, 3-hr resolution, and 50 sigma ($\sigma$) layers was produced using the Regional Ocean Modeling System (ROMS), a free-surface, hydrostatic, primitive equation model discretized with a terrain following vertical coordinate system (Shchepetkin and McWilliams, 2005). Runs initialize every 7 days and run for 10 days, covering a period from 12-November-2017 though 29-December-2017. The model is initialized using 1/12˚ resolution Operational Mercator (GLOBAL_ANALYSIS_FORECAST_PHY_001_024) and radiation/nudging lateral boundary conditions and a 3-day relaxation timescale (Marchesiello et al., 2001). Flux forcing is computed every 3 hours with turbulent fluxes calculated from bulk formulae (Fairall et al., 1996; Large & Pond, 1981) using the atmospheric state obtained from JRA-55 (Tsujino et al., 2018), and there is no imposed tidal forcing. The model uses an orthogonal curvilinear grid that tracks latitude/longitude lines. Tracers and momentum use 3rd order upstream-biased advection in the horizontal, and 4th order centered differences advection in the vertical. The model uses GLS vertical mixing parameterization (Warner et al., 2005) for turbulent mixing of momentum and tracers; with the Kantha and Clayson (1994) stability function, Craig & Banner (1994) wave breaking surface flux, and Charnok surface roughness from wind stress (Carniel et al., 2009). Horizontal diffusion of tracers and momentum were 2 $m^2/s$ and 3 $m^2/s$, with quadratic bottom friction with coefficient 0.003.

[4] This mechanism is similar to that of Gula et al. [2016] for centrifugal instability in the Gulf Stream.

Upwind advection schemes contain implicit smoothing; dynamical processes below 5-km (rather than 2-km) are not well represented due to smoothing over the stencil, staggered grids, and time stepping. A limitation of using the 1-km ROMS model to study instability is its resolution constraints; submesoscale instabilities undoubtedly exist below the scales represented in the model. While symmetrically unstable flows may persist in the ocean due to forcing (e.g., FSI), another limitation of the model is that instabilities can persist longer in the model than the ocean due to lack of removal mechanisms (either a resolved forward energy cascade, or a parameterization for the unresolved forward energy cascade). We are confident that the westward velocity anomalies (relative to geostrophic flow) along the Bank which decrease stability are not simply pressure gradient errors (Mellor et al., 1998); which would manifest as a spurious addition of ~0.01-0.1 m/s in the same direction as the geostrophic current (eastward, with the coast to the left). We also validate the results using a feature model to demonstrate the physical conditions leading to submesoscale instability (Appendix).

Vertical buoyancy ($B$) and velocity ($U$, $V$) profiles are linearly interpolated from $\sigma$-coordinates to a uniform vertical grid ($\Delta z = 5$ m) before calculation of spatial derivatives and the subsequent application of instability criteria (Table 1). The distribution of instability is examined from the perspective of meridional sections (conducive to study of the mainly-zonal PF jet), as well as the full 3-D domain. The latitude ($\phi$) and velocity of the PF jet were obtained at the location of the maximum eastward component of velocity $U(\theta, \phi, z)$ north of 56.5ºS and within the longitudinal range for which the front and associated jet are dominantly zonal, 63.0˚W $< \theta <$ 60.0˚W. The purpose of the latitudinal constraint is to avoid misidentifying the SACCF jet core or (secondary filaments associated with the PF) as the PF jet core, and a possible limitation of this method is that it neglects curvatures (relative angle) of Namuncurá - Burwood Bank and the PF jet. We use longitude 60.5˚W to illustrate meridional sections, but its features are common to meridional slices where the PF jet is principally zonal.

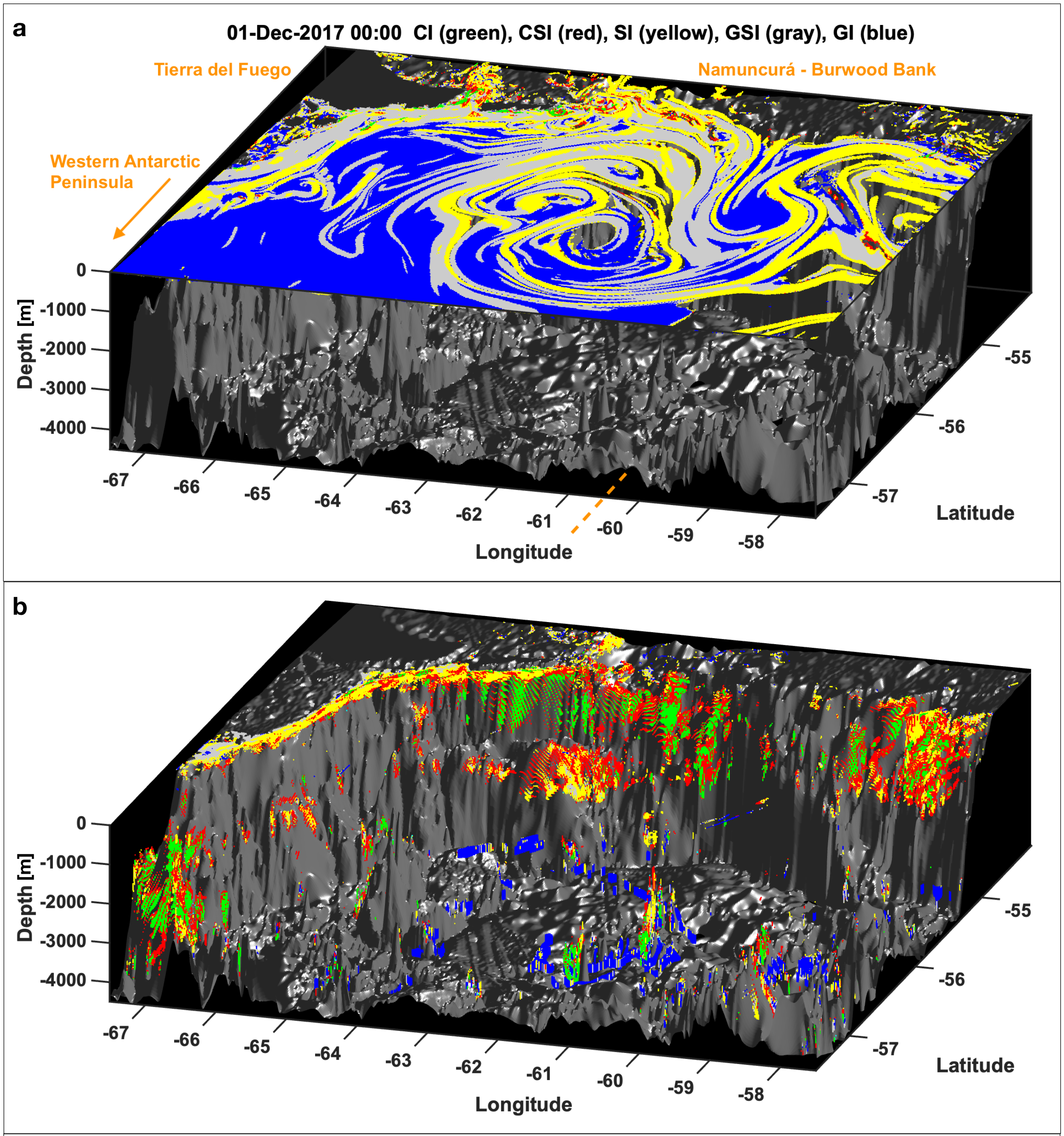

**Figure 2:** Showing unstable nodes in the **(a)** upper 100 m and **(b)** lower 100-4500 m for a timestep (01-Dec-2017) of the 3-D model domain; including gravitational (GI, blue), gravitational-symmetric (GSI, gray), centrifugal (CI, green), centrifugal-symmetric CSI, (red), and symmetric (SI, yellow) instability.

## Results

As D'Asaro et al. (2011) and Thomas et al. (2013) hypothesized, submesoscale instabilities including CI, CSI, and SI are present in the upper ocean of the ACC (Fig. 2a), both at the abrupt lateral buoyancy gradients of PF filaments and where submesoscale vortices are generated by interaction between the ACC and Tierra del Fuego and advected eastward. In the subsurface (Fig. 2b), instabilities concentrate on the north side of the zonal jet (as predicted by geostrophic instability theory), but are tied to topography; these instabilities are found where the PF jet experiences topographic drag along Namuncurá - Burwood Bank. The position of the PF jet (Fig.

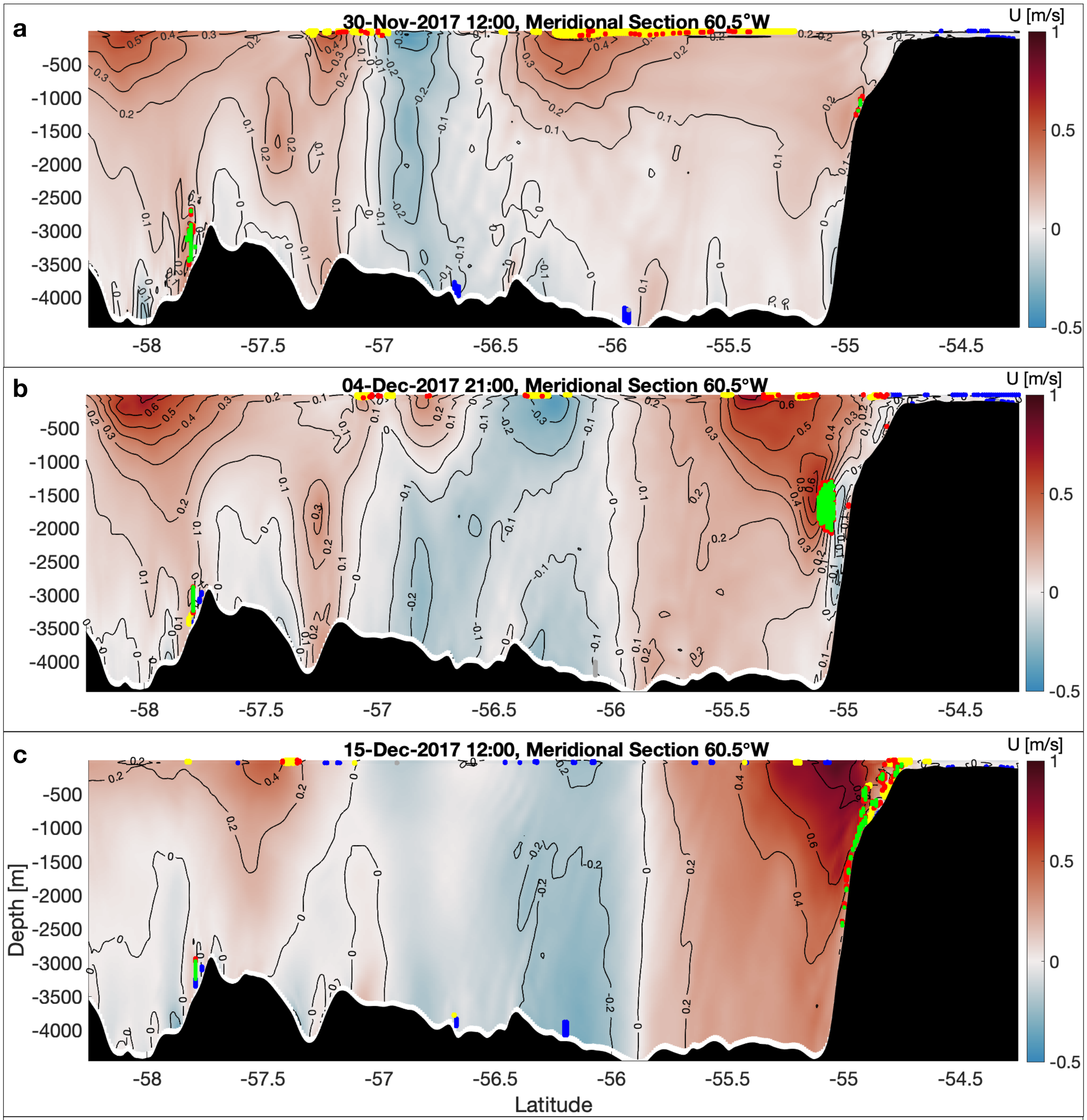

**Figure 3:** Phenomenology of centrifugal (CI, green), centrifugal-symmetric (CSI, red), symmetric (SI, yellow), gravitational-symmetric (GSI, gray), and gravitational (GI, blue) instabilities at 60.5ºW (dotted line in Fig. 2) for states of the PF jet with contours of eastward velocity (*U*). **(a)** Showing SI and CSI in the SBL, associated with Polar Front filaments. **(b-c)** Showing CSI and CI in the subsurface, associated with flow-topography interaction along Tierra del Fuego and Namuncurá - Burwood Bank.

3), namely, its proximity to the continental rise, controls the amount of subsurface CI, CSI, and SI. In general, a southern (northern) hemisphere process causing an increase (decrease) in relative vorticity reduces the potential vorticity towards becoming symmetrically unstable. Here, topographic drag on the north edge of the ACC (or alternatively, the southern flank of an abyssal feature) increases horizontal shear to create instability. The potential vorticity in Fig. 3c is decomposed into its individual terms (Eq. 1) and provided in Fig. 4. This decomposition illustrates that the low potential vorticity required for overturning instability is enduringly produced when vorticity of the fluid is spun in the anticyclonic direction (Fig. 4b). Conversely, SI at open-ocean fronts is dependent on weak stratification (see pale layer in Fig. 4a) for the production of net-positive potential vorticity (Eq. 1) and is thus confined to the weakly stratified SBL (~0-100 m).

For the subset of the domain where the PF jet is principally zonal (see Methods and box in Fig. 5a) we identify submesoscale instabilities (Fig. 5b) in relation to latitude and speed of the PF (Fig. 5c). Separating the domain into the shallow and subsurface ocean (Fig. 5d, Fig. 5e) demonstrates two distinct instability regimes: an SBL dominated by classic shear-convective instability, and the subsurface ocean --- where CI and CSI contribute more greatly

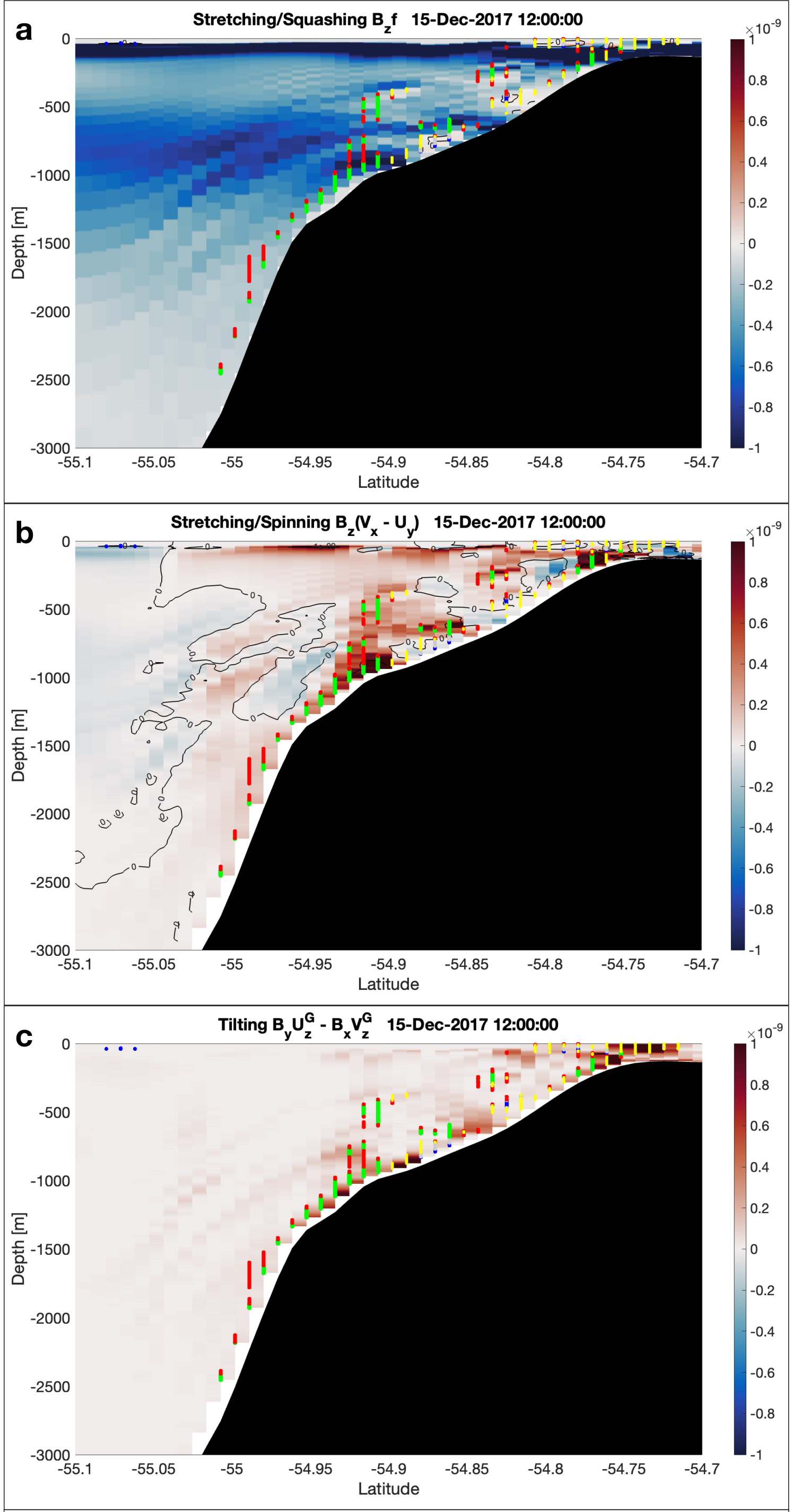

**Figure 4:** **(a)** Stretching and squashing, **(b)** stretching and spinning, and **(c)** tilting terms [$s^{-3}$] of Ertel potential vorticity ($q$, Eq. 1) for the boundary region in Fig. 3c, where $qf < 0$ is unstable. Positive values (red tones) are destabilizing, and negative values (blue tones) are stabilizing. Spinning (squashing) is the primary driver of CSI (GSI). SI arises from combined effects of squashing, spinning, and tilting. Ribbon-like features are $\sigma$-coordinates.

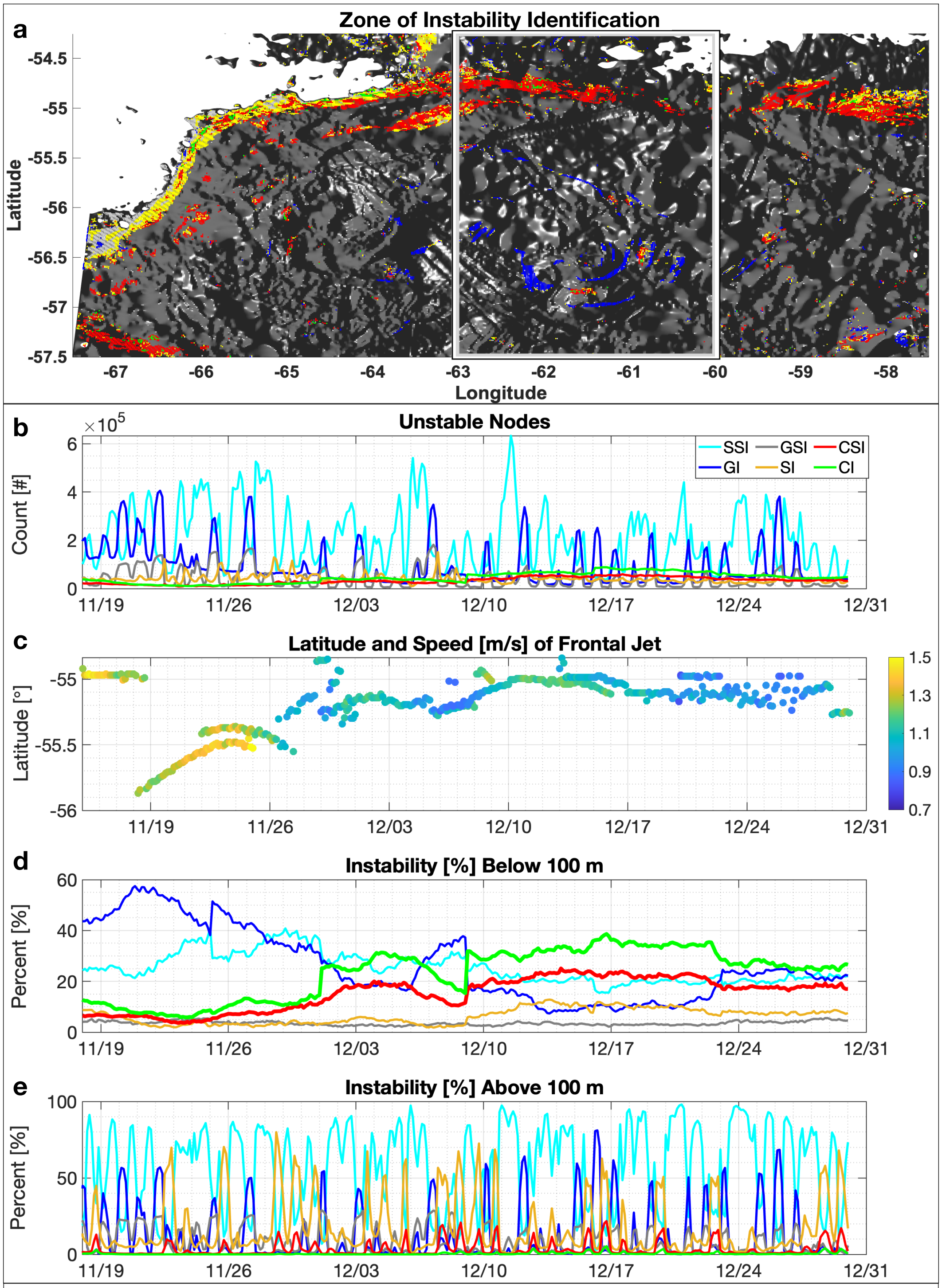

**Figure 5:** Instability compared to frontal jet variability. Showing **(a)** the region of instability identification (white box); **(b)** the number of uniformly interpolated grid nodes which are unstable to stratified shear instability (SSI), GI, GSI, SI, CSI, and CI; and **(c)** latitude and speed of the jet core. Namuncurá - Burwood Bank is located at ~55ºS. Additionally showing the relative prevalence of each instability type identified in both **(d)** the subsurface ocean and **(e)** the upper 100 m of the region.

to the instability budget.

In the subsurface ocean (Fig. 5d), the location of the PF jet and the associated mesoscale eddy present in the model (Fig. 1a, Fig. 5a - box) control the relative role of each instability types to which the flow is predisposed. While the PF jet is shifted southward in late November, CI and CSI comprise about 10% of all overturning instabilities (as defined in the Fig. 5 caption). While the PF shifts northward towards Namuncurá - Burwood Bank (~55ºS) in the beginning of December, their relative role increases to about 30% and dominates the subsurface instability budget. Conversely, GI decreases as the front and mesoscale eddy shift northward. The GI arises in the modeled abyss (Fig. 2b) when deep-reaching flow of the mesoscale eddy stirs dense bottom water equatorward to overlie lighter water; its setup depends on the proximity of the eddy to the bottom water at its poleward source. Its possible importance to abyssal mixing is beyond the scope of this paper but is worthy of future inquiry.

In the SBL (Fig. 5e), the relative prevalence of each instability type is characterized by episodic surface evolution (see Ferris et al., 2022a) as well as a diurnal oscillation produced by convective forcing in localized regions of the domain. The diurnal oscillation augments the total amount of both GI and GSI in the SBL and juxtaposes the steady nature of instability in the subsurface ocean (Fig. 5d) --- as with FSI due to winds (Dong et al., 2021), topographic drag provides a mechanism for sustained overturning instability in the subsurface ocean. It is worth underscoring that analytical criteria in Table 1 are derived for a steady flow, such that they are meaningful only if instabilities grow on a timescale faster than the timescale at which the flow evolves. Diurnal variation of SI, GI, and GSI (Fig. 5e) implies that some perceived instability in the SBL is transient and vanishes (by cessation of convective forcing and restratification) before undergoing forward energy cascade, such that some of the instability in Fig. 5e is not physically meaningful.

## Discussion

Submesoscale instability is found at two major sites in the ACC: in the weakly stratified SBL, and in the subsurface ocean near lateral boundaries (or equivalently, sloping topography). The amount of subsurface submesoscale instability arising in a ACC jets depends on their location with respect to topography, indicating that the importance of submesoscale instability to forward energy cascade in the ACC depends on temporal variation of the PF (unlike SBL instabilities which are not tied to geography of the PF). Furthermore, topographic shearing of the PF jet presents a mechanism for sustained CI, CSI, and SI (analogous to FSI in the upper ocean). Both atmospheric changes, such as the Southern Annular Mode and the El Niño Southern Oscillation (ENSO), and internal dynamical variabilities alter the position of the ACC's frontal jets on an intra-annual to inter-annual timescale (Gille et al., 2016). Altering the latitude of the ACC fronts with respect to Southern Ocean topography likely impacts the role of CI, CSI, and SI in mixing near surface and topographic boundaries (this study); as well as that of more ubiquitous internal wave processes (see Fig. 4 and Fig. 5 of St. Laurent et al., 2012; Waterhouse et al., 2014) which impact vertical heat, carbon, and nutrient flux throughout the Southern Ocean.

The mechanism for Antarctic Intermediate Water (AAIW) formation is not well known; and we speculate that enhanced mixing at the ACC's northern boundary due to the topographic shearing mechanism presented in this paper may play a role. We wonder whether some of the near-boundary elevated TKE (over rough topography and along continental margins) which was historically attributed to internal waves could be, in part, from submesoscale instabilities undergoing forward energy cascade. However, this is not the first finding of topographic shearing facilitating the forward energy cascade by producing submesoscale instabilities in a major current; Gula et al. (2016) observed a similar mechanism. CI, CSI, and SI are created on the anticyclonic side of the ACC when topographic drag increases relative vorticity enough to destabilize the flow. If presence of northern boundary controls the development of instability, a natural conclusion is that the Drake Passage and Scotia Sea region are unique to the rest of the ACC (perhaps with the exceptions of the Agulhas Bank and Campbell Plateau); however, other features such as the Kerguelen Plateau (as well as submerged seamounts in currents across the global ocean, such as the New England Seamount Chain in the Gulf Stream) provide topographic drag and are thus candidates for topographic forcing of submesoscale instability. Despite being tied to specific topographic features, evidence for the spatial inhomogeneity of Southern Ocean mixing (Tamsitt et al., 2017) shows that the spatial extent of a particular mixing mechanism does not equate to its overall impact. Topographically-sheared instabilities in the ACC may disproportionately affect mixing.

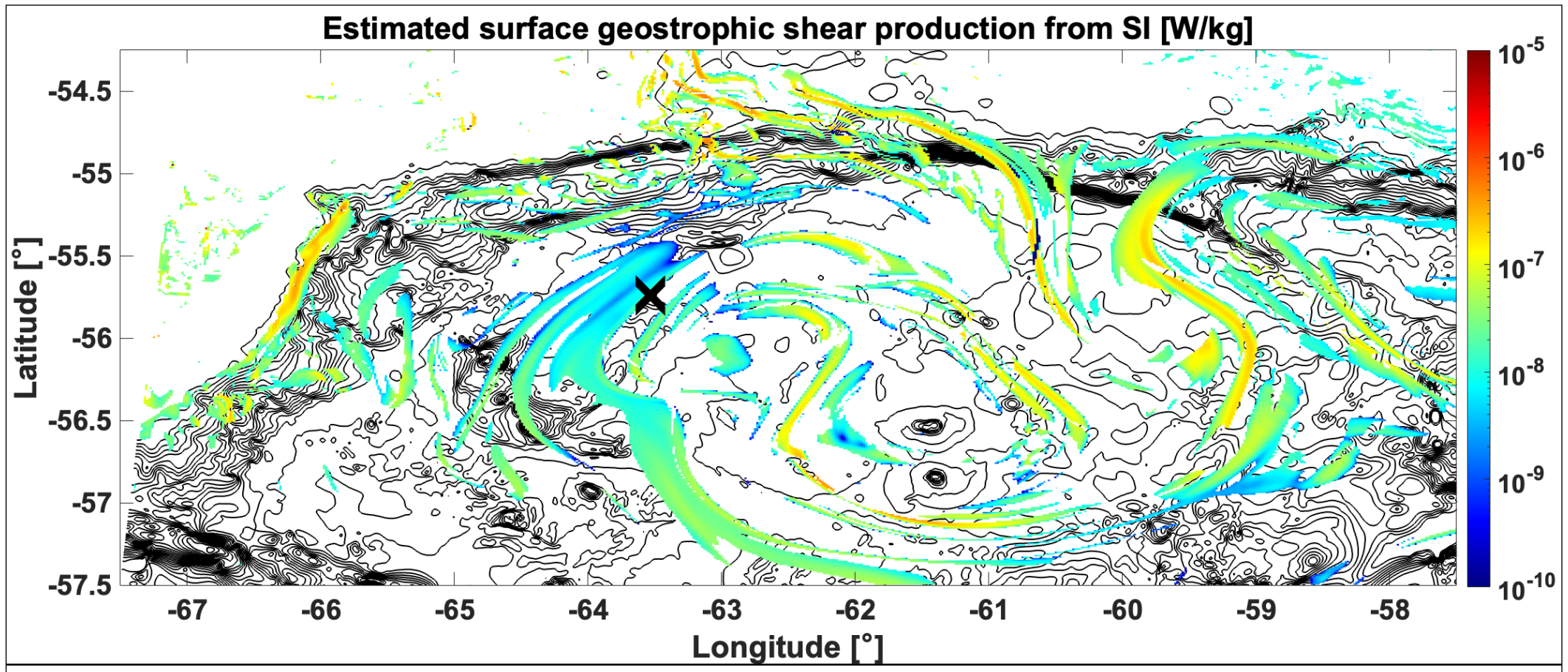

**Figure 6:** Estimated geostrophic shear production at $z = 0$ m for the timestep in Fig. 2 (01-Dec-2017), estimated after (Thomas et al., 2013; Bachman et al., 2017; Dong et al., 2021). Bathymetric contours are underlaid at 300-m intervals and the instantaneous location of the glider (Fig. 1b) is marked with a cross.

Unexplained are the physical processes responsible for the elevated TKE dissipation rates observed in the subsurface open ocean (1-12 December) of AUSSOM (Fig. 1b), during which the glider sampled the core of the PF down to 350 m. Estimates of geostrophic shear production due to SI in the SBL (Fig. 6) are indeed consistent with observed TKE dissipation rates; but SI cannot be responsible for this feature due to its limited vertical influence ($0 < z < 100$ m). The physical processes depositing TKE in the subsurface open ocean --- which we speculate to involve critical layer interactions with internal lee waves at the margins of Polar Front jets --- must be addressed in future work.

We have provided insight into the relative role and spatial arrangement of symmetric instabilities in the Southern Ocean, finding that submesoscale overturning instabilities may be as important along the topographic boundaries of the ACC as they are at fronts in the SBL of the open ocean. This finding is relevant to other energetic currents rich in frontal structure; instability analysis of the Kuroshio and Gulf Stream are not much more common than that of the ACC, and there has been little submesoscale instability work in the subarctic (which is similarly rich in energetic filaments, complex topography, and sharp density fronts). The inclusion of topographically sheared submesoscale instabilities may be important for modeling ocean structure in several coastal and littoral regions regions of the world. Meanwhile, the overall velocity structure in many of these regions is altered by tides with short periods, complicating the applicability of existing balanced frameworks.

This study and others (Dewar et al., 2015; Gula et al., 2016; Wenegrat et al., 2020; Yankovsky et al., 2021) are strong support that traditionally surface-associated submesoscale frontal instabilities can arise below the SBL when forced by topography; and if parameterized in ocean models, should be treated as more than just an SBL effect. This said, we emphasize that the presence of unstable flow does not guarantee that SI or its hybrid types will grow on a meaningful timescale or produce a TKE contribution (Ferris & Gong, 2024). An open task is to estimate the mixing efficiency associated with topographically forced submesoscale instability (Ijichi et al., 2020), and its relative role (if any) in driving upper ocean structure. The community's need to develop realistic parameterizations for unresolved submesoscale instability is strong motivation to make further observations in regions suspected of topographically-sheared SI, CSI, and CI; and to better understand the growth rates, depths, and re-stratification timescales associated with these instabilities in the real ocean. At the same time, equal focus should be placed on classic shear turbulence and internal wave phenomena which are likely as important (if not more important) than submesoscale instability away from boundaries.

**Acknowledgements**

Ferris's effort was supported by the Virginia Institute of Marine Science (VIMS) Office of Academic Studies and the Applied Physics Laboratory - University of Washington Science & Engineering Enrichment & Development Fellowship. AUSSOM microstructure data is used courtesy of Louis St. Laurent and was supported by the OCE Division of the National Science Foundation, award 1558639 (PIs St. Laurent and Sophia Merrifield). ROMS output is used courtesy of Harper Simmons and John Pender, for which computational resources were provided by Research Computing Systems at University of Alaska Fairbanks. We thank the VIMS Ocean-Atmosphere & Climate Change Research Fund (Emeritus John Boon) for his gift of a computer. We thank Justin Shapiro and Thilo Klenz for sharing technical expertise.

**Appendix**

We validate the results of the 1-km ROMS hindcast using velocity-based feature model after Gangopadhyay & Robinson (2002). A 2D idealized model (Table A1), based on a PF-associated jet observed during AUSSOM, was created from the geometry observed via AVISO, wind conditions from CCMP V2.0, and approximate density from the glider to investigate the development of instability.

| Table A1. Parameters for 2D model | |
|---|---|
| Domain height | $H = 3000$ m |
| Domain width | $L = 200$ km |
| Vertical grid divisions | $NZ = 51$ ($\Delta z \approx 60$ m) |
| Horizontal grid divisions | $NY = 301$ ($\Delta y \approx 0.67$ km) |
| Base latitude | 57°S |
| Reference density | $\rho_0 = 1027$ kg/m$^3$ |

The model is a cross-section of a geostrophic zonal jet with no time evolution. The background density structure ($\rho_\theta$) based on a Drake Passage width of $L_{DP} = 850$ km and potential density anomaly ($\delta\rho_\theta$) which produce the geostrophic jet are given by:

$$\rho_\theta(z, y) = (1 + zr_z + yr_y) + \delta\rho_\theta \quad \text{(X1a)}$$

$$\delta\rho_\theta(z, y) = -0.06(1 + z/H)\tanh((y - y_{0\rho})/\Delta y_\rho) \quad \text{(X1b)}$$

where $r_z = -1/(H\rho_0)$ and $r_y = -0.3/(L_{DP}\rho_0)$ and width of the anomaly ($\Delta y_\rho$) is 7 km, resulting in stratification $N^2 = 3 \times 10^{-6}$s$^{-2}$. The latitude of the density anomaly ($y_{0\rho}$) is chosen to be 100 km (Case Ocean, representing an open ocean jet) or 170km (Case Boundary, representing near-boundary jet). The jet velocity ($U_0 \leq 1.37$ m/s) is calculated using thermal wind balance $U_z(z, y) = -B_y/f$, where $B(z, y) = -g\rho/\rho_0$ and subscripts indicate differentiated quantities. A horizontal velocity anomaly $\delta U \geq -0.35$ m/s (Eq. X2b) and logarithmic decay function are used to represent the effects of topographic shearing (Eq. X2c). The horizontal velocity anomaly is equivalent to representing the effects of a topographic form drag using the expression for wall vorticity, $\zeta = -\tau_D/(\rho_0\nu_h) = C_D U\,|\,U\,|$.

The physical presence of topographic shearing along Namuncurá - Burwood Bank is supported by two datasets: (a) a weak westward flow 0.1-0.2 m/s was observed in the 2020 reoccupation of GO-SHIP sections SR1B and A23 across the Drake Passage and Scotia Sea (Firing, 2020), and (b) a westward velocity anomaly (intermittently amounting to a westward flow, e.g. Fig. 3b) of similar magnitude arises in our ROMS model.

$$U_0(z,y) = U(z-\Delta z, y) + U_z(z,y)\Delta z \tag{X2a}$$

$$\delta U(z,y) = -0.35\tanh((y - y_{0_U})/\Delta y_U) \tag{X2b}$$

$$U(z,y) = \begin{cases} U_0 + \delta U, & \text{if } y \le y_{0_U} \\ (U_0 + \delta U)\dfrac{\ln(L-y)}{\ln(L-y_{0_U})}, & \text{if } y_{0_U} < y \end{cases} \tag{X2c}$$

Here $y_{0_U} = 170$ km and $\Delta y_U = 3$ km. Omitting the topographic boundary layer, the transport is similar for both idealized scenarios; 30.73 Sv for Case Ocean and 30.87 Sv for Case Boundary.

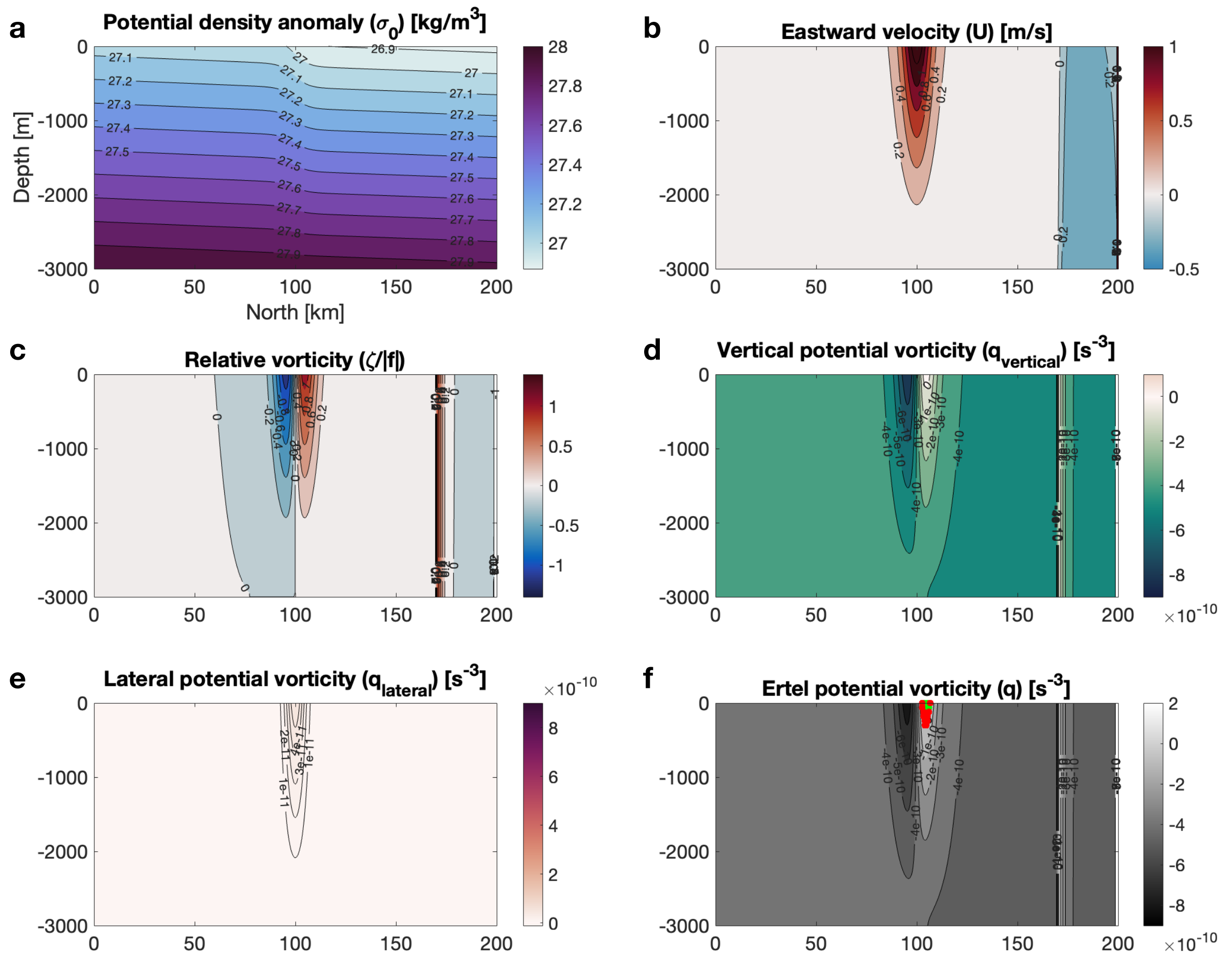

**Figure A1:** Cross-sections of an idealized jet for the near-boundary case. Nodes satisfying criteria (Table 1) for centrifugal (green) and centrifugal-symmetric instability (red) are highlighted **(f)**. There are no instances of pure symmetric instability.

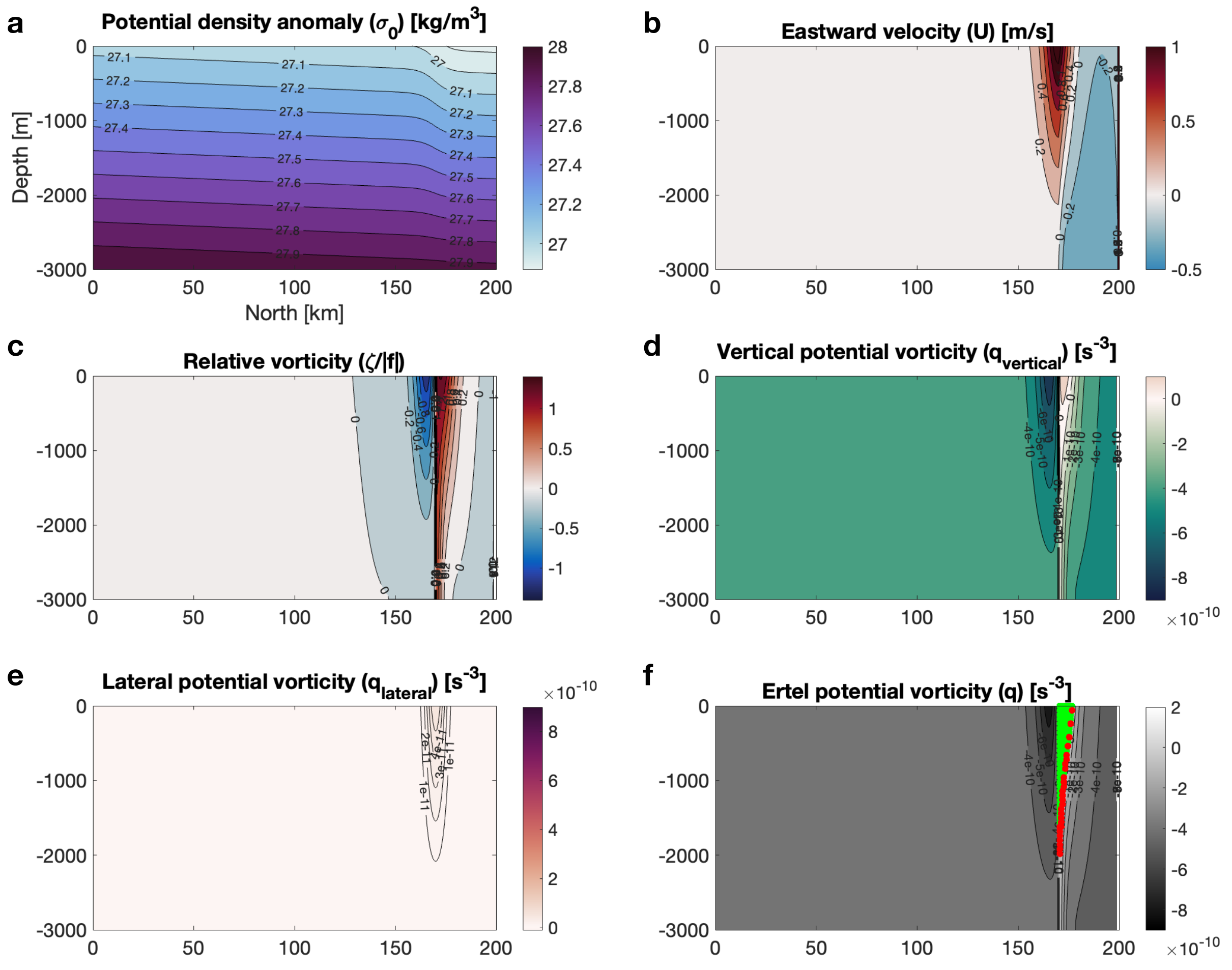

**Figure A2:** Cross-sections of an idealized jet for the near-boundary case. As in Fig. A1, nodes satisfying criteria (Table 1) for centrifugal (green) and centrifugal-symmetric instability (red) are highlighted **(f)**. There are no instances of pure symmetric instability.

Cross-sections of the jet in both cases are provided (Fig. A1 and Fig. A2) with instabilities (Table 1) highlighted over Ertel potential vorticity. In Case Ocean (Fig. A1), stratification effects create CSI which doubles the total amount of overturning instability that would otherwise be limited to CI. In Case Boundary (Fig. A2), close proximity of the jet to the northern boundary increases the instances of CI, which is augmented by a doubling in CSI. The CSI extends throughout the water column, illustrating it is not a process specific to the surface ocean as commonly intuited. This feature model validates the conclusion derived from the ROMS model, that an otherwise identical jet can produce different amounts of subsurface submesoscale instability depending on its location relative to topography.